\documentclass[aps,prb,twocolumn,groupedaddress,showpacs]{revtex4}

\usepackage{graphicx}
\usepackage{dcolumn}
\usepackage{bm}

\begin{document}

\title{Low-temperature nodal-quasiparticle transport in
lightly doped YBa$_2$Cu$_3$O$_y$ near the edge of the 
superconducting doping regime}

\author{X. F. Sun}
\email[]{ko-xfsun@criepi.denken.or.jp}
\author{Kouji Segawa}
\author{Yoichi Ando}
\email[]{ando@criepi.denken.or.jp}
\affiliation{Central Research
Institute of Electric Power Industry, Komae, Tokyo 201-8511,
Japan.}

\date{\today}

\begin{abstract}

In-plane transport properties of nonsuperconducting YBa$_2$Cu$_3$O$_y$
($y$ = 6.35) are measured using high-quality untwinned single crystals.
We find that both the $a$- and $b$-axis resistivities show $\log (1/T)$
divergence down to 80 mK, and accordingly the thermal conductivity data
indicate that the nodal quasiparticles are progressively localized with
lowering temperature. Hence, both the charge and heat transport data do
not support the existence of a ``thermal metal" in nonsuperconducting
YBa$_2$Cu$_3$O$_y$, as opposed to a recent report by Sutherland {\it et
al.} [Phys. Rev. Lett. {\bf 94}, 147004 (2005)]. Besides, the present
data demonstrate that the peculiar $\log (1/T)$ resistivity divergence
of cuprate is {\it not} a property associated with high-magnetic fields.

\end{abstract}

\pacs{74.25.Fy, 74.25.Dw, 74.72.Bk}

\maketitle

The electronic properties of underdoped cuprates have been 
extensively studied in recent years, with the aim of 
understanding how the strong electron correlations in these 
materials manifest themselves and lead to high-$T_c$ 
superconductivity.\cite{Orenstein} Most notably, it has been 
elucidated that the electronic density of states near the Fermi 
energy $E_F$ is gradually suppressed from above $T_c$ in 
underdoped cuprates, causing a ``pseudogap" in the normal 
state.\cite{Orenstein,Timusk} The pseudogap opens anisotropically 
in the Brillouin zone and has four nodes along the zone 
diagonals, similarly to the $d$-wave superconducting (SC) gap. 
Furthermore, the pseudogap is formed through partial destruction 
of the Fermi surface from the antinodes, leaving ``Fermi arcs" 
near the zone diagonals.\cite{Norman} These Fermi arcs are 
observed not only in the normal state of moderately underdoped 
superconductors but also in lightly doped nonsuperconducting 
(NSC) samples,\cite{Yoshida} which has lead to the idea that the 
Mott-insulating state of parent cuprates turns into a metal-like 
state through creation of {\it nodal quasiparticles} on the Fermi 
arcs.\cite{Yoshida,Ando_Hall} Intriguingly, the ``metallic" 
transport supported by the Fermi arcs in lightly doped cuprates 
has surprisingly good mobility at moderate 
temperatures\cite{mobility} and shows a behavior reminiscent of 
ordinary Fermi liquids ($T^2$ inelastic scattering rate and 
$T$-independent Hall coefficient).\cite{Ando_Hall} 

It should be noted, however, that those nodal quasiparticles in 
lightly doped cuprates are bound to be localized in the $T 
\rightarrow 0$ limit, giving rise to variable-range hopping (VRH) 
behavior at low temperature.\cite{mobility} The nodal 
quasiparticles in the SC state, on the other hand, are known to 
be {\it delocalized} in the $T \rightarrow 0$ limit and can carry 
heat,\cite{Durst,Taillefer} which has been documented by the 
measurements of the thermal conductivity $\kappa$ at milli-Kelvin 
temperatures that find a finite $T$-linear term expected for 
fermionic excitations throughout the superconducting doping 
regime of various 
cuprates.\cite{Takeya,Sun_YBCO,Bi2201,Sutherland} Notably, it is 
also known that the normal state of sufficiently underdoped 
cuprates becomes ``insulating" in the $T \rightarrow 0$ limit 
when the superconductivity is suppressed with a high magnetic 
field;\cite{Ando_logT,Boebinger,Fournier,Ono} in fact, the 
magnetic-field ($H$) dependence of $\kappa$ in the SC state is 
indicative of a magnetic-field-induced quasiparticle localization 
at sufficiently low doping,\cite{Sun_YBCO,Bi2201,LSCO_H} which 
seems to be consistent with the insulating normal state under 
high magnetic fields. Hence, at low doping, it appears that the 
nodal quasiparticles are delocalized only in the SC state, and 
either suppression of the superconductivity by high magnetic 
fields or underdoping into the NSC regime cause the nodal 
quasiparticles to be localized. This picture is consistent with 
the idea that there is a ``hidden" metal-insulator (MI) 
transition at a critical doping $p_c$ in the SC doping regime, 
and below $p_c$ the nodal quasiparticles are localized in the 
absence of superconductivity. 

However, Sutherland {\it et al.}\cite{Sutherland} have recently 
argued that the ground state of lightly-doped YBa$_2$Cu$_3$O$_y$ 
(YBCO) in the NSC regime is a ``thermal metal", proposing that 
the nodal quasiparticles can remain delocalized in the underdoped 
NSC regime of a clean cuprate. This claim implies that $p_c$ 
falls into the NSC doping regime in YBCO, which questions the 
universality of the hidden MI transition in the SC regime. 
Naturally, if the nodal quasiparticles indeed remain delocalized 
in a NSC sample, one should be able to confirm the metallic 
nature by measuring the resistivity, since there is no 
superconductivity that short-circuits the dc charge transport; 
unfortunately, Sutherland {\it et al.} relied only on heat 
transport measurements for drawing their 
conclusion.\cite{Sutherland} Indeed, the intrinsic resistivity 
behavior of YBCO near the edge of the SC doping regime has been 
notoriously difficult to be elucidated, mostly because the 
resistivity data are easily contaminated by filamentary 
superconductivity at such doping.

In this paper, we critically examine whether there is really a 
thermal metal phase in clean YBCO in the NSC regime, by measuring 
both the charge and heat transport properties on untwinned single 
crystals down to 70--80 mK. We have succeeded in preparing 
samples that are free from filamentary superconductivity and 
found that at $y$ = 6.35, which is very close to the SC doping 
regime ($y \ge 6.40$), the temperature dependences of the $a$- 
and $b$-axis resistivities ($\rho_a$ and $\rho_b$) are both well 
described by $\log (1/T)$ in zero field, which obviously 
indicates that the nodal quasiparticles are localized in the 
zero-temperature limit. Also, this charge-transport behavior 
signifies that the nodal quasiparticles are {\it inelastically} 
scattered down to the lowest temperature, and hence dictates that 
one should never assume the electronic thermal conductivity to be 
linear in $T$, an assumption that was employed in the analysis by 
Sutherland {\it et al.}\cite{Sutherland} but is only valid in the 
elastic scattering regime of fermionic excitations. In fact, the 
behavior of the $a$- and $b$-axis thermal conductivities 
($\kappa_a$ and $\kappa_b$) indicates that the nodal 
quasiparticles are progressively localized upon lowering 
temperature.

The high-quality single crystals of YBa$_2$Cu$_3$O$_y$ are grown 
by a flux method using Y$_2$O$_3$ crucibles to avoid inclusion of 
impurity atoms;\cite{Segawa,Segawa_Hall} we have 
documented\cite{Sun_YBCO} that our crystals are of comparable 
cleanliness to the best crystals grown in BaZrO$_3$ 
crucibles.\cite{Zhang,Hill} Using elaborate procedures to 
reliably control the oxygen content $y$ described in Ref. 
\onlinecite{Segawa_Hall}, the samples are carefully annealed and 
detwinned. Note that our samples are quenched after the 
high-temperature annealing to avoid long-range oxygen ordering, 
although formations of short-range Cu-O chain fragments always 
take place in low-doped YBCO samples at room temperature after 
quenching --- this is the cause of the ``room-temperature 
annealing" (RTA) effect.\cite{Sutherland,Segawa} The $y$ values 
of the samples reported here are 6.45, 6.40, and 6.35, for which 
the zero-resistivity $T_c$'s are 20, 7.5, and 0 K, respectively. 
The SC samples ($y$ = 6.45 and 6.40) are measured after one week 
of RTA, after which the change of $T_c$ almost saturates. On the 
other hand, the $y$ = 6.35 samples are measured within one day of 
RTA, because the one-week RTA tends to cause a filamentary 
superconductivity which contaminates the resistivity data below 2 
K; thus, the present $y$ = 6.35 samples can be considered to be 
essentially oxygen-order free. The resistivity and thermal 
conductivity are independently measured by using the conventional 
four-probe method\cite{mobility} and a steady-state 
technique,\cite{Takeya,Sun_YBCO} respectively. 

\begin{figure}
\includegraphics[clip,width=8.5cm]{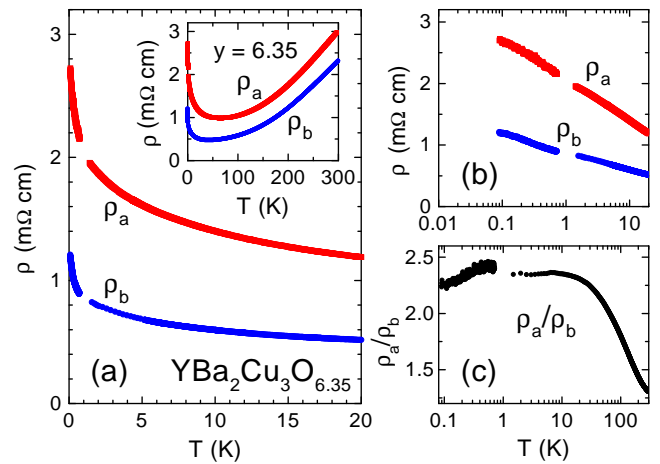}
\caption{(a) $T$-dependences of zero-field $\rho_a$ and $\rho_b$
for $y$ = 6.35 below 20 K; inset shows the data up to 300 K.
(b) Semilog plot of the data to show the $\log (1/T)$ divergence.
(c) $T$-dependence of the in-plane anisotropy, $\rho_a/\rho_b$.
The slight discontinuity in the data across 1 K is due to 
the use of a very small excitation current in the 
dilution-fridge measurements.}
\end{figure}

The inset to Fig. 1(a) shows the temperature dependences of 
$\rho_a$ and $\rho_b$ for $y$=6.35 below 300 K, where one can see 
that the resistivity shows a metallic behavior down to 70 K (40 
K) along the $a$ ($b$) axis, but a steep upturn sets in at lower 
temperature. The main panel of Fig. 1(a) zooms in on the data 
below 20 K; here, one can clearly see that both $\rho_a$ and 
$\rho_b$ are apparently diverging at low temperature and it is 
difficult to assert some definite resistivity value for their $T 
\rightarrow 0$ limit. In fact, as shown in Fig. 1(b), the $T$ 
dependences of both $\rho_a$ and $\rho_b$ are well described by 
$\log (1/T)$ below $\sim$20 K down to 80 mK (more than two 
decades), and the $\log (1/T)$ dependence mathematically suggests 
that the resistivity grows infinitely large for $T \rightarrow 
0$. Note that this behavior is the same as that found in the 
normal state of underdoped cuprate superconductors under high 
magnetic fields,\cite{Ando_logT,Ono} but this is the first time 
that a clear $\log (1/T)$ behavior is observed in zero magnetic 
field: When $T_c$ is suppressed to zero by Zn doping in 
underdoped samples, the resistivity shows a divergence quicker 
than $\log (1/T)$.\cite{Hanaki_Zn,Komiya_Zn} Incidentally, we 
found that application of a 16-T magnetic field causes negligible 
change in the resistivity data of the present samples. Hence, the 
present data demonstrate that the $\log (1/T)$ resistivity 
divergence of cuprates is {\it not} a property associated with 
high-magnetic fields; also, the $\log (1/T)$ behavior in clean 
YBCO suggests that this peculiar behavior is not merely a result 
of disorder-induced localization but is fundamentally related to 
the strong-correlation physics.

Figure 1(c) shows the $T$ dependence of the in-plane resistivity
anisotropy, $\rho_a/\rho_b$, for $y$=6.35. As we have reported
previously,\cite{anisotropy} $\rho_a/\rho_b$ grows to as large as 
$\sim$2.5 at this doping, even though the orthorhombicity is very 
weak (0.2\%) and is about to disappear --- this is arguably the 
most convincing evidence that the electrons self-organize into a 
macroscopically anisotropic state (such as a nematic 
charge-stripe phase\cite{Kivelson_RMP}) in lightly-doped 
YBCO.\cite{anisotropy} In Fig. 1(c), one can see that 
$\rho_a/\rho_b$ becomes nearly $T$-independent when the 
resistivity is showing the $\log (1/T)$ divergence. This is 
rather reminiscent of the behavior of the out-of-plane 
anisotropy, $\rho_c/\rho_{ab}$, in underdoped LSCO, which also 
shows a saturation when the resistivities individually show the 
$\log (1/T)$ divergence.\cite{Ando_logT}

\begin{figure}
\includegraphics[clip,width=8.5cm]{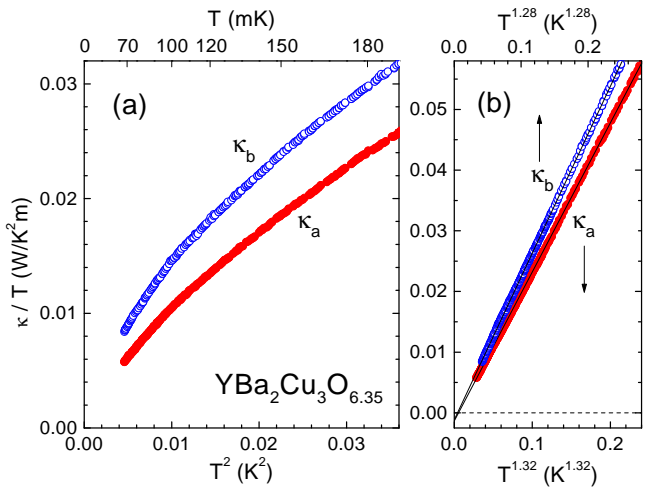}
\caption{(a) $T$-dependences of $\kappa_a$ and $\kappa_b$ for $y$ = 6.35
down to 70 mK, plotted in $\kappa/T$ vs $T^2$.
(b) $\kappa/T$-vs-$T^{\alpha}$ analysis of $\kappa_a$ and $\kappa_b$ for
$y$ = 6.35 to artificially bring out a linear behavior.}
\end{figure}

Figure 2 shows the corresponding $\kappa_a$ and $\kappa_b$ data 
for $y$=6.35. In Fig. 2(a), $\kappa/T$ is plotted versus $T^2$, 
which would allow one to separate the electronic contribution 
$\kappa_e$ ($\sim T$ in the elastic-scattering regime) from the 
phononic contribution $\kappa_p$ ($\sim T^3$ in the 
boundary-scattering regime), provided that the data at the lowest 
temperatures [usually below 100--200 mK (Refs. 
\onlinecite{Taillefer,Takeya,Sun_YBCO,Bi2201,Behnia})] fall onto 
a straight line; however, one can easily see in Fig. 2(a) that 
both the $\kappa_a$ and $\kappa_b$ data are continuously curved 
in this plot down to our lowest temperature, 70 mK, and one 
cannot make a linear fitting for any reasonably extended 
temperature range. In fact, the data look as if $\kappa/T$ is 
heading to zero at $T$=0. Note that, by choosing some fractional 
power of $T$ for the horizontal axis, one could make a plot that 
shows an almost linear behavior [Fig. 2(b)], and this fact 
corroborates the observation that the data are nowhere linear in 
the ordinary $\kappa/T$ vs $T^2$ plot. Actually, as we discuss 
later, when the resistivity shows a $\log (1/T)$ divergence and 
hence an inelastic scattering process is affecting the electron 
transport, there is no reason to believe that the electronic part 
of $\kappa/T$ is constant, and therefore it is not meaningful to 
play around with the data to try to separate $\kappa_p$ by making 
a plot like Fig. 2(b). In other words, the absence of the linear 
behavior in the $\kappa/T$ vs $T^2$ plot in the present case does 
{\it not} mean that the boundary-scattering regime of phonons 
(where $\kappa_p \sim T^3$) is never achieved nor the proper 
power for $\kappa_p$ is different from 3,\cite{Sutherland} but 
it means that $\kappa_e$ never behaves as $\sim T$ due to the 
$\log (1/T)$ localization behavior. We note that in our 
experiments both the resistivity and thermal conductivity results 
for $y$=6.35 are confirmed to be essentially reproduced in at 
least one more set of samples.

\begin{figure}
\includegraphics[clip,width=8.5cm]{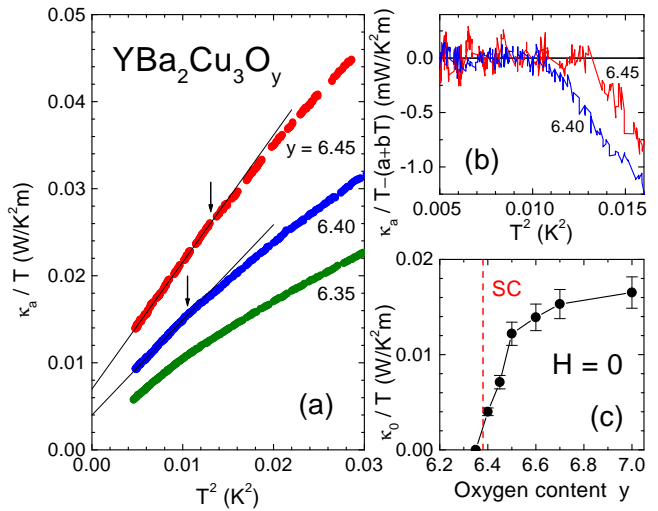}
\caption{(a) $\kappa_a/T$ vs $T^2$ plot for $y$ = 6.45, 6.40, 
and 6.35; solid lines are the straight-line fits to the 
lowest-temperature data of SC samples, and arrows mark the temperature 
below which the fitting is good.
(b) Difference between the $\kappa_a/T$ data and the linear 
fittings shown in (a) for the two dopings.
(c) Doping dependence of $\kappa_0/T$ at $H$=0 determined in the 
present work and in Ref. \onlinecite{Sun_YBCO}; note that all the data 
are for the $a$-axis (perpendicular to the chains).}
\end{figure}

It should be remarked that there is a notable difference in the thermal
conductivity behavior between SC samples and NSC samples; namely, for SC
samples there is always a finite range of temperature where the data are
linear in the $\kappa/T$ vs $T^2$ plot with a finite intercept at $T$=0,
as shown in Fig. 3(a), while for NSC samples the data in such a plot are
curved all the way down to the lowest $T$. To make it easier to judge
the rationality of the linear fittings for the SC samples, Fig. 3(b)
shows the difference between the data and the fits; here, one can see
that the linear fitting is very reasonable below 0.013 and 0.010 K$^2$
({\it i.e.}, 115 and 100 mK) for $y$ = 6.45 and 6.40, respectively.
Physically, the observed change in the thermal conductivity behavior
across the superconductivity boundary is a manifestation of the fact
that the nodal quasiparticles are localized on the NSC side of this
boundary, while they are delocalized in the SC state.

In passing, we note that the linear fittings in Fig. 3(a) for $y$ = 6.45
and 6.40 are very consistent with what is expected for $\kappa_p$: In
the boundary-scattering regime, $\kappa_p$ is given by $\frac{1}{3}\beta
\langle v_{ph} \rangle l_{ph} T^3$, with $\beta$ the phonon specific
heat coefficient, $\langle v_{ph} \rangle$ the averaged sound velocity,
and $l_{ph}$ the phonon mean-free path; in perfect crystals $l_{ph}$
takes the maximum value $1.12\bar{w}$ with $\bar{w}$ the geometric mean
width of the sample. Using the parameters for YBCO cited in Ref.
\onlinecite{Taillefer} and $\bar{w}$ of 0.356 (0.245) mm for $y$ = 6.45
(6.40), we obtain $l_{ph}/1.12\bar{w}$ of 0.85 (0.92).

Let us now discuss the most important question of whether the 
ground state at $y$=6.35 is a metal or an insulator. For this 
discussion, it is crucial to examine the meaning of the $\log 
(1/T)$ resistivity divergence. Although this behavior is not 
completely equivalent to the $\log T$ behavior of the 
conductivity $\sigma$ in 2D weak localization,\cite{Lee_RMP} it 
is instructive to recall the basic physics behind the weak 
localization. According to the scaling theory of weak 
localization,\cite{Lee_RMP} the inelastic diffusion length $L_i$ 
(which grows upon lowering $T$) sets the length scale to 
determine the size-dependent conductivity $g(L_i)$, which depends 
logarithmically on $L_i$ and vanishes for $L_i \rightarrow 
\infty$; this dependence on $L_i$ is the essential source of the 
$\log T$ behavior.\cite{note} Therefore, when the $\log T$ weak 
localization behavior is observed, the transport is governed by 
the inelastic scattering process and one can never assume 
$\kappa_e$ to behave as $\sim T$ as in the elastic-scattering 
regime. In fact, since all the electrons are eventually localized 
in the $\log T$ regime, $\kappa_e/T$ is bound to vanish for $T 
\rightarrow 0$.

In the case of the $\log (1/T)$ resistivity behavior in cuprates,
although the exact mechanism to produce the logarithmic dependence would
be different from the weak localization, the system is certainly not in
the elastic-scattering regime and some inelastic process continues to be
effective for $T \rightarrow 0$, causing the ground state at $T$=0 to be
a correlated insulator. This means that it makes best sense to interpret
our data for $y$=6.35 to be indicative of a progressive localization of
electrons and vanishing $\kappa_e/T$ for $T \rightarrow 0$. Also, the
$\log (1/T)$ resistivity behavior implies that it is not reasonable to
assert $\kappa_e/T$ to be constant and hence the
$\kappa/T$-vs-$T^{\alpha}$ analysis like that in Fig. 2(b) is
essentially meaningless. In this context, it is also pointless to
examine the validity of the Wiedemann-Franz (WF) law in the $\log
(1/T)$-insulating cuprates, because the WF law holds only in the
elastic-scattering regime.\cite{Ashcroft}

Having discussed that the data for $y$=6.35 are indicative of a
vanishing $\kappa_e/T$ for $T \rightarrow 0$, we can now draw a 
diagram for the doping dependence of the $T$=0 residual 
electronic term, $\kappa_0/T$, as shown in Fig. 3(c), where the 
data obtained in our previous work\cite{Sun_YBCO} for $6.50 \le y 
\le 7.00$ are also plotted. Altogether, there is a good 
systematics in the doping dependence of $\kappa_0/T$, which is 
essentially consistent with that for LSCO,\cite{Takeya} giving 
confidence that the simultaneous disappearance of the 
superconductivity and delocalized nodal quasiparticles is a 
fundamental nature of the cuprates. The present result also 
indicates that even in the cleanest cuprate YBCO the ground state 
of the lightly-doped NSC regime is an insulator, though the 
nature of the insulating state, which is characterized by the 
$\log (1/T)$ resistivity divergence here, is different from that 
in lightly doped LSCO where it is due to the disorder-driven 
localization.\cite{mobility}

Lastly, let us briefly discuss the possibility that, while the charges
are localized and cannot give rise to a thermal metal phase in the NSC
regime of YBCO, some exotic fermionic excitations might be alternatively
responsible for a thermal metal in the charge insulating state. One can
easily see that this scenario is very unlikely, because, as shown in
Fig. 3(a), $\kappa_a/T$ decreases notably and systematically from $y$ =
6.45 to 6.35, which speaks against the idea that some new fermions start
to carry heat for $y \le 6.35$ and contribute a sizable $\kappa_0/T$.

In conclusion, we find that in clean, untwinned single crystals of
nonsuperconducting (NSC) YBCO at $y$ = 6.35, the temperature dependences
of both $\rho_a$ and $\rho_b$ show $\log (1/T)$ divergence in zero
magnetic field, which indicates that the nodal quasiparticles in the NSC
regime are localized for $T \rightarrow 0$ and that they are
inelastically scattered down to the lowest temperature. The latter means
that the electronic thermal conductivity $\kappa_e$ cannot be linear in
$T$ even at the lowest temperature, and accordingly the data for
$\kappa_a$ and $\kappa_b$ at $y$ = 6.35 are indicative of a
$T$-dependent $\kappa_e/T$ that appears to vanish for $T \rightarrow 0$.
Hence, the critical examination of the heat transport at $y$ = 6.35 in
the light of the charge-transport behavior leads to a conclusion that
the ground state of clean underdoped cuprate in the NSC regime is {\it
not} a thermal metal but is a peculiar insulator characterized by the
$\log (1/T)$ resistivity divergence.


We thank Louis Taillefer for useful discussions and sharing preprints
prior to publication. This work was supported by the Grant-in-Aid for
Science provided by the Japanese Society for the Promotion of Science.



\begin{thebibliography}{}

\bibitem{Orenstein}
J. Orenstein and A. J. Millis, Science {\bf 288}, 468 (2000).

\bibitem{Timusk}
T. Timusk and B. Statt, Rep. Prog. Phys. {\bf 62}, 61 (1999).

\bibitem{Norman}
M. R. Norman {\it et al.}, Nature (London) {\bf 392}, 157 (1998).

\bibitem{Yoshida}
T. Yoshida {\it et al.}, Phys. Rev. Lett. {\bf 91}, 027001 (2003).

\bibitem{Ando_Hall}
Y. Ando {\it et al.}, Phys. Rev. Lett. {\bf 92}, 197001 (2004).

\bibitem{mobility}
Y. Ando {\it et al.}, Phys. Rev. Lett. {\bf 87}, 017001 (2001).

\bibitem{Durst}
A. C. Durst and P. A. Lee, Phys. Rev. B {\bf 62}, 1270 (2000).

\bibitem{Taillefer}
L. Taillefer {\it et al.}, Phys. Rev. Lett. {\bf 79}, 483 (1997).

\bibitem{Takeya}
J. Takeya, Y. Ando, S. Komiya, and X. F. Sun, Phys. Rev. Lett.
{\bf 88}, 077001 (2002).

\bibitem{Sun_YBCO}
X. F. Sun, K. Segawa, and Y. Ando, Phys. Rev. Lett. {\bf 93},
107001 (2004).

\bibitem{Bi2201}
Y. Ando {\it et al.}, Phys. Rev. Lett. {\bf 92}, 247004 (2004).

\bibitem{Sutherland}
M. Sutherland {\it et al.}, Phys. Rev. Lett. {\bf 94}, 147004 (2005).

\bibitem{Ando_logT}
Y. Ando {\it et al.}, Phys. Rev. Lett. {\bf 75}, 4662 (1995).

\bibitem{Boebinger}
G. S. Boebinger {\it et al.}, Phys. Rev. Lett. {\bf 77}, 5417
(1996).

\bibitem{Fournier}
P. Fournier {\it et al.}, Phys. Rev. Lett. {\bf 81}, 4720 (1998).

\bibitem{Ono}
S. Ono {\it et al.}, Phys. Rev. Lett. {\bf 85}, 638 (2000).

\bibitem{LSCO_H}
X. F. Sun, S. Komiya, J. Takeya, and Y. Ando, Phys. Rev. Lett.
{\bf 90}, 117004 (2003); D. G. Hawthorn {\it et al.}, Phys. Rev. 
Lett. {\bf 90}, 197004 (2003). 

\bibitem{Segawa}
K. Segawa and Y. Ando, Phys. Rev. Lett. {\bf 86}, 4907 (2001).

\bibitem{Segawa_Hall}
K. Segawa and Y. Ando, Phys. Rev. B {\bf 69}, 104521 (2004).

\bibitem{Zhang}
Y. Zhang {\it et al.}, Phys. Rev. Lett. {\bf 86}, 890 (2001).

\bibitem{Hill}
R. W. Hill {\it et al.}, Phys. Rev. Lett. {\bf 92}, 027001 (2004).

\bibitem{Hanaki_Zn}
Y. Hanaki, Y. Ando, S. Ono and J. Takeya, Phys. Rev. B {\bf 64},
172514 (2001).

\bibitem{Komiya_Zn}
S. Komiya and Y. Ando, Phys. Rev. B {\bf 70}, 060503(R) (2004).

\bibitem{anisotropy}
Y. Ando, K. Segawa, S. Komiya, and A. N. Lavrov, Phys. Rev. Lett.
{\bf 88}, 137005 (2002).

\bibitem{Kivelson_RMP}
S. A. Kivelson {\it et al.}, Rev. Mod. Phys. {\bf 75}, 1201
(2003).

\bibitem{Behnia}
S. Nakamae {\it et al.}, Phys. Rev. B, {\bf 63}, 184509 (2001).

\bibitem{Lee_RMP}
P. A. Lee and T. V. Ramakrishnan, Rev. Mod. Phys. {\bf 57}, 287
(1985).

\bibitem{note}
The $\log T$ behavior of weak localization should saturate at low enough
$T$ in a finite system, where the sample size sets the upper limit for
$L_i$; in our data in Fig. 1(b), the slight saturation tendency
discernible at the lowest $T$ could be a similar finite-size effect, and
it does {\it not} automatically suggest a conceptually metallic ground
state. 

\bibitem{Ashcroft}
N. W. Ashcroft and N. D. Mermin, {\it Solid State Physics}
(Holt-Saunders, Philadelphia, 1976), p. 323.

\end{thebibliography}
\end{document}